\documentclass{ws-ijmpd}

\let\trimmarks\nocropmarks
\usepackage{epsfig,amssymb}
\newcommand{\beq}{\begin{equation}}
\newcommand{\eeq}{\end{equation}}
\newcommand{\bqa}{\begin{eqnarray}}
\newcommand{\eqa}{\end{eqnarray}}
\newcommand{\fr}{\frac}

\newcommand{\lb}{\label}

\begin{document}
\title{A NOTE ON THE CYLINDRICAL COLLAPSE OF COUNTER-ROTATING DUST}
\author{\sc S\'{e}rgio M. C. V. Gon\c{c}alves\footnote{Corresponding author. E-mail address: sergio.goncalves@yale.edu}}
\address{Department of Physics, Yale University, New Haven, Connecticut 06511, U.S.A. \\
Theoretical Astrophysics, California Institute of Technology, Pasadena, California 91125, U.S.A.}
\author{\sc Sanjay Jhingan}
\address{Yukawa Institute for Theoretical Physics, Kyoto University, Kyoto 606-8502, Japan}
\maketitle
\pub{Received (received date)}{Revised (revised date)}
\begin{abstract}
We find analytical solutions describing the collapse of an infinitely long cylindrical shell of counter-rotating dust. We show that---for the classes of solutions discussed herein---from regular initial data a curvature singularity inevitably develops, and {\em no} apparent horizons form, thus in accord with the spirit of the hoop conjecture.
\end{abstract}

\section{Introduction}

One of the outstanding issues of general relativity is that of the final state of gravitational collapse. Understandably, much has been written about spherical collapse~\cite{review}, but comparatively less is known about non-spherical geometries. Thorne's seminal analysis of cylindrical collapse \cite{thorne65} led him to formulate the {\em hoop conjecture}, which essentially states that horizons form when and only when the gravitational mass of the system is confined within a maximum ``circumference'' in every direction~\cite{thorne72}. Subsequent numerical analyses of prolate and oblate collapse~\cite{shapiro&teukolsky91-92}, and of gravitational radiation emission in aspherical collapse~\cite{nakamura&shibata&nakao93} were unable to refute Thorne's conjecture.

One of the best studied non-spherical cases is that of an infinite cylindrical shell of counter-rotating dust. A detailed analysis by Apostolatos and Thorne~\cite{apostolatos&thorne92}, using $C$-energy~\cite{thorne65b} balance arguments for sequences of momentarily static radiation free configurations, showed that collapse is halted by an infinitesimal amount of counter-rotation (and very likely, also by infinitesimally small net rotation); no horizons, {\em nor} curvature singularities form. A subsequent, more detailed analysis, of the non-rotating dust case, confirmed the non-occurrence of horizons~\cite{echeverria93}.

All of the analyses mentioned above resorted to balance arguments and/or analytical approximations for the late stages of collapse, which is highly relativistic, and can thus be well modeled by the collapse of a null fluid. As far as we can ascertain, the first analytical, exact, fully dynamical solution is the one given recently by Pereira and Wang (PW), who studied the case of an infinite cylindrical shell of counter-rotating dust, with a Minkowski interior and an outgoing null fluid exterior~\cite{pereira&wang00}. By resorting to one simplifying ansatz, PW obtained analytical solutions that describe the implosion of the shell, leading to a central curvature singularity. PW presented these solutions, but did not analyze~\cite{pw} the following aspects, which are crucial for the issue of cosmic censorship: (i) regularity of (and inherent constraints on) the initial data; (ii) existence, or otherwise, of trapped surfaces in the spacetime, and in particular on the initial spacelike slice; (iii) satisfaction of energy conditions by the stress-energy tensor on the shell. In this paper, we perform such an analysis, thereby establishing the regularity of the initial data and physical reasonability conditions. Our results show that generic regular initial data can be specified, there are no trapped surfaces in the spacetime, and a curvature line-like singularity forms. This is in accord with the spirit of the hoop conjecture, but provides an example of a naked (at least locally) singularity.

Geometrized units, in which $8\pi G=c=1$, are used throughout.

\section{Cylindrical thin shell of counter-rotating dust}

The model consists of an infinite cylindrical thin shell $\Sigma$, joining a Minkowski interior (with coordinates $\{x_{-}^{\mu}\}=\{t,r,z,\phi\}$), to an exterior spacetime metric (with coordinates $\{x_{+}^{\nu}\}=\{T,R,z,\phi\}$) describing an outgoing null fluid:
\bqa
ds^{2}_{-}&=&dt^{2}-dr^{2}-dz^{2}-r^{2}d\phi^{2}, \lb{gin} \\
ds^{2}_{+}&=&e^{-b(\xi)}(dT^{2}-dR^{2})-dz^{2}-R^{2}d\phi^{2}, \lb{gout}
\eqa
where $\xi=T-R$. The exterior stress-energy tensor is $T_{\mu\nu}^{+}=(db/d\xi)R^{-1}k_{\mu}k_{\nu}$, where $k_{\mu}$ is the generator of outgoing radial null geodesics:
\beq
k_{\mu}=\fr{1}{\sqrt{2}}(1,-1,0,0).
\eeq
On $\Sigma$, where intrinsic coordinates $\{\xi^{a}\}=\{\tau,z,\phi\}$ are defined, the 3-metric $\gamma_{ab}$ is obtained by continuity of the interior and exterior metrics across $\Sigma$, giving:
\bqa
ds_{\Sigma}^{+}&=&d\tau^{2}-dz^{2}-R^{2}_{0}(\tau)d\phi^{2}, \lb{gshell} \\
d\tau&=&\sqrt{1-r'^{2}_{0}}dt=e^{-b(\xi_{0})/2}\sqrt{1-R'^{2}_{0}}dT,
\eqa
where the prime denotes ordinary differentiation with respect to the argument; $r_{0}(t)=R_{0}(T)$ defines the shell, and $\xi_{0}\equiv T-R_{0}(T)$. It then follows that
\beq
\left(\fr{dt}{dT}\right)^{2}=R'^{2}_{0}+e^{-b(\xi_{0})}(1-R'^{2}_{0})\equiv\Delta.
\eeq

The extrinsic curvature, $K_{ab}^{\pm}$, is given by~\cite{israel66}
\beq
K_{ab}=\fr{\partial x^{\alpha}}{\partial \xi^{a}}\fr{\partial x^{\beta}}{\partial \xi^{b}}
\nabla_{\alpha}n_{\beta}=-n_{\nu}\left(\fr{\partial^{2}x^{\nu}}{\partial\xi^{a}\partial\xi^{b}}+\Gamma^{\nu}_{\alpha\beta}\fr{\partial x^{\alpha}}{\partial \xi^{a}}
\fr{\partial x^{\beta}}{\partial \xi^{b}}\right),
\eeq
where $n_{\mu}$ is the unit-normal to $\Sigma$. On the shell, the field equations reduce to the Darmois-Israel junction conditions,
\beq
S_{ab}=[K_{ab}]-\gamma_{ab}[K],
\eeq
where $[K_{ab}]\equiv K_{ab}^{+}-K_{ab}^{-}$, $[K]\equiv\gamma^{ab}[K_{ab}]$, and $S_{ab}$ is the stress-energy tensor on the shell:
\beq
S_{ab}=\rho\delta^{\tau}_{a}\delta^{\tau}_{b}+p_{z}\delta^{z}_{a}\delta^{z}_{b}+p_{\phi}\delta^{\phi}_{a}\delta^{\phi}_{b},
\eeq
where $\rho$, $p_{z}$, and $p_{\phi}$ are the surface density, and principal pressures, respectively, and are given by~\cite{fnote}
\bqa
\rho&=&\fr{e^{b/2}}{R_{0}\Delta^{1/2}\sqrt{1-R'^{2}_{0}}}\left[
(\Delta+R'^{2}_{0})-\Delta^{1/2}({1+R'^{2}_{0})} \right], \lb{dens} \\
p_{\phi}&=&\fr{e^{b/2}}{\Delta^{1/2}(1-R'^{2}_{0})^{3/2}}\left[
R''_{0}(\Delta^{1/2}-1)+\fr{b'}{2}(1-R'_{0}) \right. \\ \nonumber
&&\left.\times(1-R'^2_{0})(\Delta^{1/2}-R'_0)\right], \lb{ptan} \\
p_{z}&=&p_{\phi}-\rho.
\eqa
Since all the Darmois-Israel junction conditions and field equations were used to derive these equations, they constitute a complete set. Following~\cite{apostolatos&thorne92}, we set $p_{z}=0\,\Rightarrow\,\rho=p_{\phi}$, which gives:
\beq
R''_{0}=(1-R'^{2}_{0})
\left[\fr{\Delta^{1/2}-R'^{2}_{0}}{R_{0}}-\fr{b'(\Delta^{1/2}-R'_0)}{2(\Delta^{1/2}-1)}(1-R'_{0})
\right].
\eeq
We now solve this equation by resorting to a similar ansatz of that used in~\cite{pereira&wang00}, which consists in isolating the $b(\xi_{0})$ dependent part, solving first for $R_{0}(T)$, and then for $b(\xi_{0})$. In what follows, we shall consider two classes of solutions.

\subsection{Class I solutions}

Denoting
\beq
\beta:=\fr{\Delta^{1/2}-R'^{2}_{0}}{R_{0}}-\fr{b'(\Delta^{1/2}-R'_0)}{2(\Delta^{1/2}-1)}(
1-R'_{0}), \label{bet}
\eeq
we set $\beta=\mbox{const.}$, and look for solutions of the equation
\beq
R''_{0}=\beta(1-R'^{2}_{0}). \label{dyn}
\eeq
Once we obtain $R_{0}(T)$, we can then solve Eq. (\ref{bet}) for $b$, thereby completing the solution. 

Initial data consists of two arbitrary functions, giving the initial velocity profile $R'_{0}(0)$, and the initial radius $R_{0}(0)$ of the shell. Provided $R_{0}(T)\in C^{3}$, the metric is regular (i.e., at least $C^{2}$) on $\Sigma_{T=0}$, and, by virtue of Eq. (\ref{dyn}), it will also be regular in the domain of dependence of the initial slice. As we shall see below, $R_{0}(T)\in C^{\infty}([0,+\infty))$, and thus generic regular initial data is always specifiable. We remark that, for the adopted $\beta=\mbox{const.}$ ansatz, the dynamics is uniquely determined from Eq. (\ref{dyn}), for a given value of $\beta$. However, to show that the field equations admit a complete self-consistent solution, one must also prove that Eq. (\ref{bet}) admits a unique analytic solution, for given initial data. This is explicitly checked below.

Equation (\ref{dyn}) has the general solution
\beq
R_{0}(T)=\fr{1}{\beta}\ln[\cosh(c_{1})\cosh(\beta T)+\sinh(c_{1})\sinh(\beta T)]+c_{2},
\lb{sol}
\eeq
where $c_{1}$ and $c_{2}$ are integration constants, which are fixed by the
initial data for the shell, via
\bqa
R_{0}(0)&=&\fr{1}{\beta}\ln(\cosh c_{1})+c_{2}, \\
R'_{0}(0)&=&-\beta^{2}\tanh(c_{1}). 
\eqa
Without loss of generality, we consider here the case of time-symmetric initial data: $R'_{0}(0)=0$. This implies
\beq
c_{1}=0,\;\;R_{0}(0)=c_{2},\;\;R''_{0}(0)=\beta<0,
\eeq
where the last inequality is required for implosion. The relevant equations are then
\bqa
R_{0}(T)&=&\fr{1}{\beta}\ln(\cosh(\beta T))+c_{2}, \\
R'_{0}(T)&=&\tanh(\beta T). \lb{Rpr}
\eqa
A curvature singularity will form at $T=T_{\rm c}$, given by $R(T_{\rm c})=0$:
\beq
T_{\rm c}=\fr{1}{\beta}\ln(1-\sqrt{1-e^{2c_{2}\beta}})-c_{2},
\eeq
which is positive definite, since $\beta c_{2}<0$.

Having obtained $R_{0}(T)$ in closed form, we now turn to the existence and uniqueness of the solution to Eq. (\ref{bet}). This equation is of the form $b'(T-R)=f(b,R(T))$. On the shell, this becomes $b'(T-R_{0}(T))=f(b,R_{0}(T))$. We now use the identity $db/dT=b'\times(1-R'_{0})$, to rewrite Eq. (\ref{bet}) in the canonical form 
\beq
\fr{db}{dT}=(1-R'_{0})f(b,R_{0}(T))=F(b,T), \label{canon}
\eeq
where the explicit $R_{0}$ dependence has been absorbed in $F$, upon substitution for $R_{0}(T)$. From the theory of ordinary differential equations, the solution of the first-order ODE $db/dT=F(b,T)$ exists and it is unique provided $F\in C^{0}$ and $(\partial F/\partial b)\in C^{0}$ ~\cite{odeth}. In our case, we have
\bqa
F(b,T)&=&-2\zeta\left[\beta-\fr{\chi-\tanh^{2}(\beta T)}{\beta^{-1}\ln(\cosh(\beta T))+c_{2}}\right], \\
\chi&\equiv&\sqrt{\tanh^{2}(\beta T)+e^{-b}/\cosh^{2}(\beta T)}, \\
\zeta&\equiv&\fr{\chi-1}{\chi-\tanh(\beta T)}.
\eqa
This function has a singular point at $T=T_{\rm s}=\beta^{-1}\ln(e^{c_{2}\beta}\pm\sqrt{e^{2c_{2}\beta}-1})$, which is complex, since $c_{2}\beta<0$; hence, $F\in C^{0}\, ,\,\forall\, T\in{\mathbb R}$, as desired. Now, taking the  partial derivative of $F$ with respect to $b$, we have:
\beq
\fr{\partial F}{\partial b}=-\fr{\zeta e^{-b}/\cosh^{2}(\beta T)}{\chi[\beta^{-1}\ln(\cosh(\beta T))+c_{2}]}-\fr{bF}{2}\left(\fr{1}{\chi}-\fr{\zeta}{\chi-1}\right).
\eeq
For arbitrary $T$, this function has a singular point at
\beq
b=b_{\rm s}=-\ln\left(-\fr{1}{4}[e^{2\beta T}-1]^{2}\right) +2\beta T,
\eeq
which is undefined, since the logarithmic argument is manifestly negative; hence, $F$ is $C^{1}$ with respect to $b$, as desired. Therefore, Eq. (\ref{bet}) admits a unique analytic solution.

We now turn to the issue of physical reasonability of the initial data. The weak energy condition (WEC) requires $\rho>0$ and $\rho+p_{\phi}>0$. From Eqs. (\ref{dens}), (\ref{ptan}), and (\ref{dyn}), this implies
\beq
b<0.
\eeq
Since $\rho=p_{\phi}$, the WEC implies the strong energy condition (SEC), whereby the requirement $p_{\phi}>0$ is automatically satisfied. By numerically solving Eq. (\ref{canon}), one can show that, for given initial data, $\{\beta,c_{2}\}$, it is {\em always} possible to choose $b(0)<0$, such that $b(T)<0,\forall T\in[0,T_{\rm c}]$. An explicit example is given below.

\begin{figure}
\begin{center}
\epsfysize=10pc
\epsfxsize=19pc
\epsffile{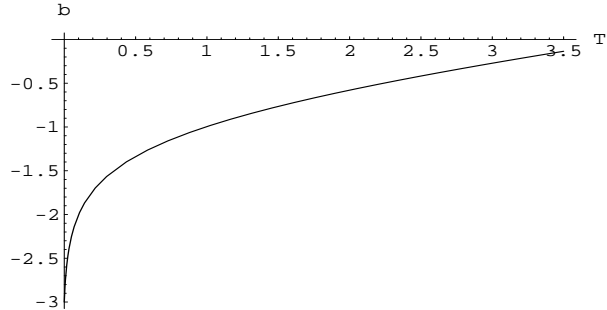}
\end{center}
\caption{Time evolution of the function $b(T-R_{0}(T))$, for the initial data set $\{\beta=-1/2, c_{2}=2, b(0)=-3\}$. For these data, $T_{\rm c}=3.31491$, and $b(T_{\rm c})=-0.184426$. \label{fig1}}
\end{figure}

Let us now examine the existence of trapped surfaces. The interior metric is flat, so it cannot contain trapped surfaces~\cite{geons}. We look for trapped surfaces in the spacetime exterior to the shell, and for $\Sigma$ itself; we note that the absence of trapped surfaces in the exterior geometry does not guarantee the same will hold for the shell, since the expansion of null geodesic congruences, $\Theta:=\nabla_{\mu}k^{\mu}$, may have a (vanishing) minimum on $\Sigma$. From Eqs. (\ref{gout}), (\ref{gshell}), we obtain
\beq
\Theta=\frac{e^{b(\xi)}}{R}[R_{0}'\delta(R-R_{0})+1],
\label{Theta}
\eeq
where $\delta(x-x_{0})$ is a Dirac-delta. It immediately follows that in the exterior spacetime $\Theta>0$, and therefore trapped surfaces can never form. On the shell, the condition for trapped surfaces is $R'_{0}=-1$. From Eq. (\ref{Rpr}), we have $R'_{0}\in(-1,0]$, which implies that the shell will never become trapped either.

Summarizing, we have obtained a class of solutions which admit generic regular initial data, are free of trapped surfaces, obey the WEC and SEC, and describe the implosion of an infinite cylindrical shell to a line-like singularity, given by the set of events $(T_{\rm c},0,z,\phi)$.

\subsection{Class II solutions}

We now define
\beq
\beta:=\left[{\Delta^{1/2}-R'^{2}_{0}}-\fr{b'(\Delta^{1/2}-R'_0)}{2(\Delta^{1/2}-1)}R'_0(1-R'_{
0})\right]. \label{bet2}
\eeq
Taking $\beta=\mbox{const.}$, we then solve for
\beq
R_{0}''=\fr{1-R_{0}'^2}{R_{0}}\beta, \label{sol2}
\eeq
which can be rewritten as
\beq
z'' =\beta(\beta + 1) z^{(\beta-1)/(\beta+1)}, \label{newsol}
\eeq
where $R_{0}\equiv z^{1/(1+\beta)}$.  Again, we shall consider time-symmetric initial data,
$R_{0}'(T=0)=0\Rightarrow z'(T=0)=0$. For implosion we need $R_{0}''(T=0)<0$ which implies $z''<0$ and $\beta>-1$, or $z''>0$ and $\beta<-1$. From Eq. (\ref{newsol}) it follows that the latter condition cannot be satisfied, and hence we must have $z''(T=0)<0$ and $-1<\beta<0$.
For general $\beta$, the solution of Eq. (\ref{newsol})  is given in terms of transcendental functions, but classes of simple analytical solutions can be found for specific values of $\beta$. As an example, we consider here the case $\beta=-1/2$. This leads to the solution
\beq
R_{0}(T) = \frac{1}{4C_1}[1-4C_1^2(T+C_2)^2], \label{sol2a}
\eeq
where $C_1$ and $C_2$ are constants of integration. The choice of time-symmetric initial data fixes $C_2=0$, which gives
\beq
R_{0}(T)=\frac{1-4C_1^2T^2}{4C_1}. \label{sol2b}
\eeq
The singularity forms at
\beq
T_{\rm c}=\frac{1}{2C_1}=2R_{0}(0) . \label{Sing}
\eeq
As expected, the time for formation of the singularity is an increasing function of the initial size of the shell.

The function $b$ is obtained from Eqs. (\ref{bet2}) and (\ref{sol2b}), via:
\beq
b'=\frac{4C_1(2\Delta^{1/2}+1-8C_1^2T^2)(\Delta^{1/2}-1)}{(\Delta^{1/2}-2C_1T)(1-4C_1^2T^
2)^2}.
\label{b}
\eeq
A similar analysis to the one for class I shows that the solution to this first-order ODE for $b$ exists and is unique. As before, both the WEC and SEC are satisfied provided $b<0$.

The equation governing trapped surfaces is, using Eqs.  (\ref{Theta}) and (\ref{sol2b}),
\beq
\Theta = \frac{4C_1 e^{b(\xi)}}{1+2C_1T},
\eeq
which is positive definite for $T>0$; there are no trapped surfaces in the spacetime.

Class II solutions constitute another example of spacetimes that admit regular initial data, no trapped surfaces, obey the WEC and SEC, and lead to a curvature singularity in a finite amount of time $T=T_{\rm c}$.

\section{Concluding remarks}
We have explicitly obtained analytical solutions describing the implosion of an infinite cylindrical shell of counter-rotating dust, which results in the formation of a line-like curvature singularity, without the formation of apparent horizons. These results provide thus further evidence in favor of the hoop conjecture, but are markedly different from previous well-known results for cylindrical counter-rotating dust collapse~\cite{thorne72,apostolatos&thorne92}, which show that collapse is halted by an arbitrarily small amount of counter-rotation, whereby the formation of a curvature singularity is prevented. We speculate that this qualitative difference is due to the matter model adopted for the interior and exterior spacetimes, since the metric on the shell---and hence its dynamics---is uniquely determined from the metrics on the matching spacetimes. In~\cite{apostolatos&thorne92}, a vacuum interior and exterior, containing cylindrical waves, was adopted, whereas the model considered here contains a flat interior and a null fluid exterior. 

We point out that, whereas the outgoing null fluid is a reasonable model for the {\em mass loss} due to the emission of massless particles and/or gravitational waves at the {\em late} stages of collapse, it fails to provide a realistic description for the {\em actual} gravitational radiation emission that results from the inward accelerated motion of the shell, since it contains an {\em ad hoc} functional form for the energy flux (modeled by the function $b$), which does not arise dynamically from the time evolution of $\Sigma$. It is therefore conceivable that a more realistic model, wherein the emission of gravitational radiation is computed directly from the shell motion (e.g., by considering a vacuum spacetime with Einstein-Rosen waves), will yield the results of~\cite{apostolatos&thorne92}, which were obtained based on balance arguments for sequences of momentarily static radiation free (MSRF) configurations~\cite{thorne01}. This expectation has been recently backed up by the results of~\cite{goncalves02}. A fully dynamical evolution~\cite{alvi&goncalves&jhingan}, and an analysis of horizon formation via global methods~\cite{goncalves&moncrief}, which do not resort to sequences of MSRF configurations nor $C$-energy balance arguments, are currently being attempted.

\section*{Acknowledgments}
It is a pleasure to thank K. Thorne for valuable discussions and comments. SMCVG acknowledges the support of FCT (Portugal) Grant PRAXIS XXI-BPD-163301-98, and NSF Grants AST-9731698 and PHY-0099568. SJ scknowledges the support of Grant-in-Aid for JSPS Fellows No. 00273.


\begin{thebibliography}{99}
\bibitem{pw}
It was not the purpose of their paper, whose main goal was to provide a general treatment for an arbitrary dynamic timelike cylindrical shell soldering two arbitrary cylindrical spacetimes.
\bibitem{fnote}
The first two equations differ slightly from the corresponding ones in~\cite{pereira&wang00}, which seem to contain an algebraic mistake; however, as we shall see below, the qualitative nature of the final solution remains unchanged.
\end{thebibliography}
\end{document}